\documentclass[aps,english,twocolumn,preprintnumbers,amsmath,amssymb,prl]{revtex4-1}
\usepackage[latin9]{inputenc}
\usepackage{amsmath}
\usepackage{amssymb}
\usepackage{graphicx}

\usepackage[T1]{fontenc}
\usepackage[latin9]{inputenc}
\usepackage{babel}

\makeatletter

\usepackage{epsfig}

\usepackage{dcolumn}

\makeatother

\begin{document}

\title{Analysis of quantum effects inside spherical charged black holes}

\author{Assaf Lanir, Amos Ori, Noa Zilberman, Orr Sela, Ahron Maline, and
Adam Levi}

\affiliation{Department of Physics, Technion, Haifa 32000, Israel}

\date{\today}
\begin{abstract}
We numerically compute the renormalized expectation value $\langle\hat{\Phi}^{2}\rangle_{ren}$
of a minimally-coupled massless quantum scalar field in the interior
of a four-dimensional Reissner-Nordstrom black hole, in both the Hartle-Hawking
and Unruh states. To this end we use a recently developed mode-sum
renormalization scheme based on covariant point splitting. In both
quantum states, $\langle\hat{\Phi}^{2}\rangle_{ren}$ is found to
approach a \emph{finite}  value at the inner horizon (IH). The final
approach to the IH asymptotic value  is marked by an inverse-power
tail $r_{*}^{-n}$, where $r_{*}$ is the Regge-Wheeler ``tortoise
coordinate'', and with $n=2$ for the Hartle-Hawking state and $n=3$
for the Unruh state. We also report here the results of an analytical
computation of these inverse-power tails of $\langle\hat{\Phi}^{2}\rangle_{ren}$
near the IH. Our numerical results show very good agreement with this
analytical derivation (for both the power index and the tail amplitude),
in both quantum states. Finally, from this asymptotic behavior of
$\langle\hat{\Phi}^{2}\rangle_{ren}$ we analytically compute the
leading-order asymptotic behavior of the trace $\langle\hat{T}_{\mu}^{\mu}\rangle_{ren}$
of the renormalized stress-energy tensor at the IH. In both quantum
states this quantity is found to diverge like $b(r-r_{-})^{-1}r_{*}^{-n-2}$
(with $n$ specified above, and with a known parameter $b$). To the
best of our knowledge, this is the first fully-quantitative derivation
of the asymptotic behavior of these renormalized quantities at the
inner horizon of a four-dimensional Reissner-Nordstrom black hole.
\end{abstract}
\maketitle
Einstein's field equations admit black hole (BH) solutions endowed
with remarkable exotic features including naked singularities, bridges
to other universes and closed timelike curves. Among these solutions
there is the Reissner-Nordstrom (RN) spacetime, describing a spherically-symmetric
BH carrying electric charge. This spacetime metric is given by{\small{}
\begin{equation}
ds^{2}=-f(r)dt^{2}+\frac{1}{f(r)}dr^{2}+r^{2}\left(d\theta^{2}+\sin^{2}\theta d\varphi^{2}\right),
\end{equation}
}where $f\left(r\right)=1-2M/r+Q^{2}/r^{2}$, $M$ and $Q$ being
respectively the mass and charge of the BH. The event horizon (EH)
and the inner horizon (IH) are respectively located at $r=r_{+}$
and $r=r_{-}$, the two solutions of $f\left(r\right)=0${\small{}
}given by $r_{\pm}=M\pm\left(M^{2}-Q^{2}\right)^{1/2}${\small{}.}
Interestingly, this metric may be analytically continued through the
BH interior into a concatenation of asymptotically flat spacetime
regions ('other universes'), accessible to an observer in the 'universe'
where the BH originally formed only by traveling through the BH. Along
the way through the BH and into the 'other universes', the observer
must cross the IH that lies inside the BH. It is a treacherous path,
however, as classical perturbations appear to form a null curvature
singularity along the Cauchy horizon (CH; the ingoing section of the
IH). This is the situation in spherically-symmetric charged BHs \cite{Hiscock:1981,PoissonIsrael:1990,OriMassInflation:1991,BradySmith:1995,Piran,Burko:1997}
as well as in spinning ones \cite{Ori:1992,BradyDrozMorsnik:1998,Ori:1999}.
Nevertheless, this null singularity, caused by classical perturbing
fields, is known to be weak \cite{Tipler} (\textit{i.e.} tidally
nondestructive \cite{Ori-weak}, with a $C^{0}$ limiting metric)
\textemdash{} in both the charged \cite{OriMassInflation:1991} and
spinning \cite{Ori:1992,Dafermos:2017} cases. 

However, a general indication that emerges from a collection of analytical
studies \cite{BirrellDavies:1978,Ottewill:2000,Hiscock:1980} on the
effect of quantum perturbations inside BHs has been that semiclassical
stress-energy fluxes are likely to diverge at the CH, although so
far it remained inconclusive in four dimensions. It is the goal of
this work to address this issue via concrete numerical calculation
(augmented by some analytical results) of the actual strength and
form of these quantum effects inside a charged BH.

Semiclassical gravity considers quantum matter fields propagating
in a classical curved spacetime. The presence of curvature ``deforms
the vacuum'' and induces a non-trivial stress energy in the quantum
fields (even in ``vacuum states''). In turn, this stress-energy
tensor deforms the spacetime metric. This back-reaction  effect is
to be determined from the semiclassical Einstein's field equation
\begin{equation}
G_{\mu\nu}=8\pi\,\langle\hat{T}_{\mu\nu}\rangle_{ren}\,.\label{semi_ein}
\end{equation}
Here $G_{\mu\nu}$ is the Einstein tensor of spacetime, and $\langle\hat{T}_{\mu\nu}\rangle_{ren}$
is the renormalized stress-energy tensor (RSET) associated with the
quantum fields.

For simplicity, our choice for a quantum field is that of a minimally-coupled
\cite{Foot1} massless scalar field, satisfying the massless Klein-Gordon
equation $\square\hat{\Phi}=0$,  where $\hat{\Phi}$ is the scalar
field operator, and $\square$ denotes the covariant D'Alembertian.
 It proves useful to first compute the renormalized vacuum expectation
value  $\langle\hat{\Phi}^{2}\rangle_{ren}$ (often called the ``vacuum
polarization''), as it is simpler than the RSET, but still captures
many of its essential features and provides important insight into
the physical content of different vacua. Furthermore, as will be seen
below, the behavior of $\langle\hat{\Phi}^{2}\rangle_{ren}$ actually
determines the divergence rate of the RSET trace $\langle\hat{T}_{\mu}^{\mu}\rangle_{ren}$
at the IH.

Semiclassical gravity predicts the evaporation of BHs through the
emission of Hawking radiation \cite{Hawking:1974,Hawking:1975}. BH
evaporation obviously implies drastic differences in spacetime structure
as compared to the corresponding classical picture.  Likewise, it
is conceivable that semiclassical stress-energy fluxes might affect
the near-CH geometry inside RN (as well  as Kerr) BHs  more strongly
than the classical perturbations  do \textemdash{} potentially converting
the CH  into a strong (i.e. tidally destructive) spacelike singularity
(and thereby preventing passage through the BH into the 'other universes').
However, these issues remained unresolved and to address them one
must, obviously, compute the RSET in the interior region of BHs, and
especially  near the CH. We have therefore set out to ultimately
compute the RSET in BH interiors, and we present here novel results
for a first step in this direction: the numerical computation of $\langle\hat{\Phi}^{2}\rangle_{ren}$
throughout the interior region \cite{Foot2} of a RN BH \cite{Foot1a},
followed by  analysis of the leading-order behavior of $\langle\hat{\Phi}^{2}\rangle_{ren}$
and also $\langle\hat{T}_{\mu}^{\mu}\rangle_{ren}$ near the CH.

The renormalization of the divergent $\langle\hat{\Phi}^{2}\rangle$
 was carried out here by the recently developed pragmatic mode-sum
(PMR) method \cite{AAt:2015,AAtheta:2016}, which numerically implements
the point-splitting renormalization scheme developed by Christensen
\cite{Christensen:1976,Christensen:1978}. This prescription for $\langle\hat{\Phi}^{2}\rangle_{ren}$
(and the same concept holds for $\langle\hat{T}_{\mu\nu}\rangle_{ren}$
as well) is depicted in the following equation: 
\begin{equation}
\left\langle \hat{\Phi}^{2}\left(x\right)\right\rangle _{ren}=\lim_{x'\rightarrow x}\left[\left\langle \hat{\Phi}\left(x\right)\hat{\Phi}\left(x'\right)\right\rangle -G_{DS}\left(x,x'\right)\right],\label{point_split}
\end{equation}
where $G_{DS}\left(x,x'\right)$ is the DeWitt-Schwinger counterterm
(explicitly given in \cite{Ander_His_Sam:1995}). 

We consider here $\langle\hat{\Phi}^{2}\rangle_{ren}$  in two quantum
vacua. One is the \textit{Unruh state} describing an evaporation of
a BH \cite{Unruh:1976}, and the other is the \textit{Hartle-Hawking
state} (HH) describing a BH in thermal equilibrium \cite{HH:1976,Israel:1976}
with an infinite bath of radiation.  In Ref. \cite{Group:2018} we
derived an explicit expression for the scalar field two-point function
in the RN interior, in both the Unruh and HH states, in terms of a
radial function $\psi_{\omega l}(r)$ which can be computed numerically.
This radial function satisfies the \emph{radial equation}: 
\begin{equation}
\frac{d^{2}\psi_{\omega l}}{dr_{*}^{2}}+\left[\omega^{2}-V_{l}\left(r\right)\right]\psi_{\omega l}=0\,,\label{radeq}
\end{equation}
where $\omega$ denotes the mode's temporal frequency and $l$ its
angular-momentum number. Here the effective potential $V_{l}(r)$
is given by 
\begin{equation}
V_{l}\left(r\right)=f(r)\left[\frac{l\left(l+1\right)}{r^{2}}+\frac{2M}{r^{3}}-\frac{2Q^{2}}{r^{4}}\right]\,,\label{v}
\end{equation}
and $r_{*}$ is the tortoise coordinate defined by  $dr/dr_{*}=f(r)$.
Note that $r_{*}\rightarrow-\infty$ ($+\infty$) at the EH (IH).
 The boundary condition for $\psi_{\omega l}$ at the EH is 
\begin{equation}
\psi_{\omega l}\cong e^{-i\omega r_{*}}\,,\,\,\,\,\,\,r_{*}\rightarrow-\infty\,.\label{initcon}
\end{equation}

The required input for the computation of $\langle\hat{\Phi}^{2}\rangle_{ren}$
inside the BH is the radial function $\psi_{\omega l}(r)$ and also
$\rho_{\omega l}^{\mathrm{up}}$, namely the reflection coefficient
for the ``up'' modes (see e.g. \cite{Group:2018}) outside the BH.
 We compute $\psi_{\omega l}(r)$ and $\rho_{\omega l}^{\mathrm{up}}$
numerically, and use them to construct the mode contributions to the
two-point function inside the BH, as prescribed in Ref. \cite{Group:2018}.
Then we regularize the mode sum using the $\theta$-splitting variant
of our method, as described in \cite{AAtheta:2016}. This same method
was implemented recently for computing $\langle\hat{\Phi}^{2}\rangle_{ren}$
inside a Schwarzschild BH in Ref. \cite{SchAssaf:2018}, where a more
detailed account of the procedure is provided. (Additional details
are provided in the Supplementary materials.)

From the symmetries of the RN geometry it immediately follows that
$\langle\hat{\Phi}^{2}\rangle_{ren}$ (like $\langle\hat{T}_{\mu}^{\mu}\rangle_{ren}$)
only depends on $r$. In the next section we present the results for
$\langle\hat{\Phi}^{2}(r)\rangle_{ren}$ throughout the range $r_{-}\leq r\leq r_{+}$.
Interestingly, it turns out that for both the Unruh and HH states,
$\langle\hat{\Phi}^{2}\rangle_{ren}$  remains finite upon approaching
the IH (although its gradient diverges there). Then subsequently we
present analytical results for the asymptotic behaviors of $\langle\hat{\Phi}^{2}\rangle_{ren}$
and $\langle\hat{T}_{\mu}^{\mu}\rangle_{ren}$ very close to the IH,
and for $\langle\hat{\Phi}^{2}\rangle_{ren}$ we also compare our
analytical and numerical results. 

\paragraph*{Numerical results: }

\begin{figure}[h!]
\centering \includegraphics[width=0.45\textwidth,height=0.28\textheight]{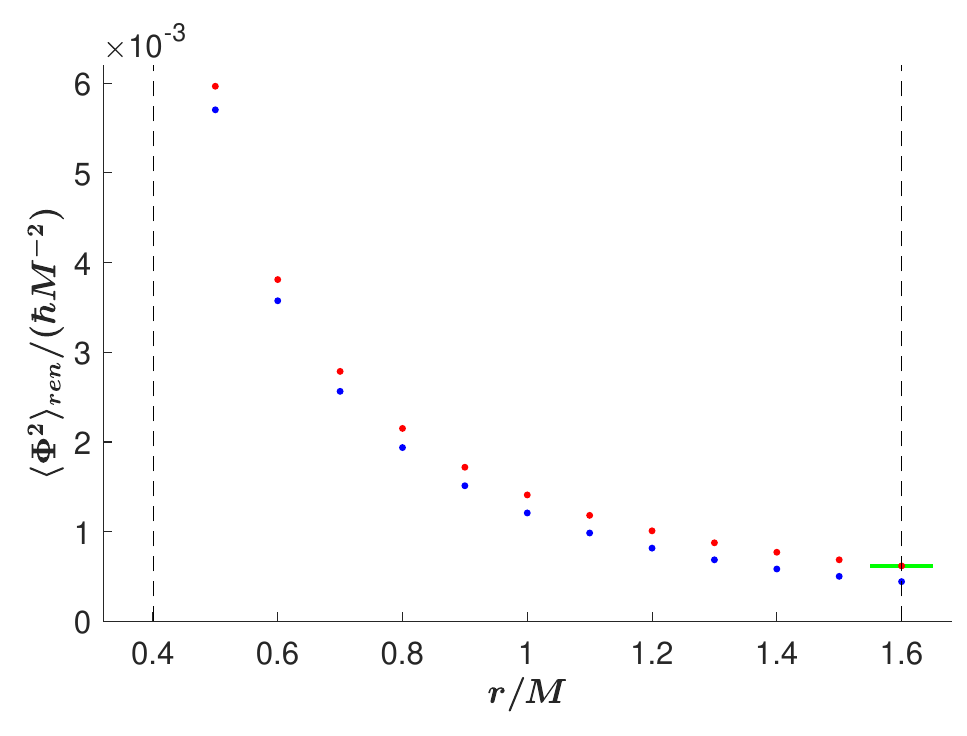}
\caption{The numerically computed $\langle\hat{\Phi}^{2}\left(r\right)\rangle_{ren}$
in the HH (red) and Unruh (blue) states in the region between the
two horizons. The short horizontal green line represents the analytical
result for $\langle\hat{\Phi}^{2}\left(r\right)\rangle_{ren}$ in
the HH state at the EH. }
\label{generalregion} 
\end{figure}
We shall focus here on the specific example $Q/M=0.8$. In this case
$r_{+}=1.6M$ and $r_{-}=0.4M$. The radial equation \eqref{radeq}
together with the initial condition \eqref{initcon} was solved numerically
for $\psi_{\omega l}(r)$, from the EH to very close to the IH, for
a sufficiently dense set of $\omega l$ modes in the range $0\leq l\leq10$
and $0<\omega<10/M$. The reflection coefficient $\rho_{\omega l}^{\mathrm{up}}$
was also computed numerically for these $\omega l$ modes. These quantities
were then used to construct $\langle\hat{\Phi}^{2}\rangle_{ren}$.
See supplementary materials for more details.

Figure 1 displays the numerical results for $\langle\hat{\Phi}^{2}\left(r\right)\rangle_{ren}$
in the region between the two horizons (specifically for $0.5\leq r/M\leq1.6$),
for both quantum states. Our result for the HH state agrees very nicely
with the known analytical result \cite{Frolov,Ottewill} at the EH,
with a difference of only $\sim0.005\%$.  

The most obvious feature seen in this figure is the steady growth
with decreasing $r$, which becomes steeper when getting close to
the IH. This trend of sharp increase towards the IH continues all
the way up to, say, $r-r_{-}\sim10^{-6}M$. From this behavior one
might get the impression (as we originally did) that $\langle\hat{\Phi}^{2}\left(r\right)\rangle_{ren}$
would diverge at the IH.

\begin{figure}[h!]
\centering \includegraphics[width=0.45\textwidth,height=0.28\textheight]{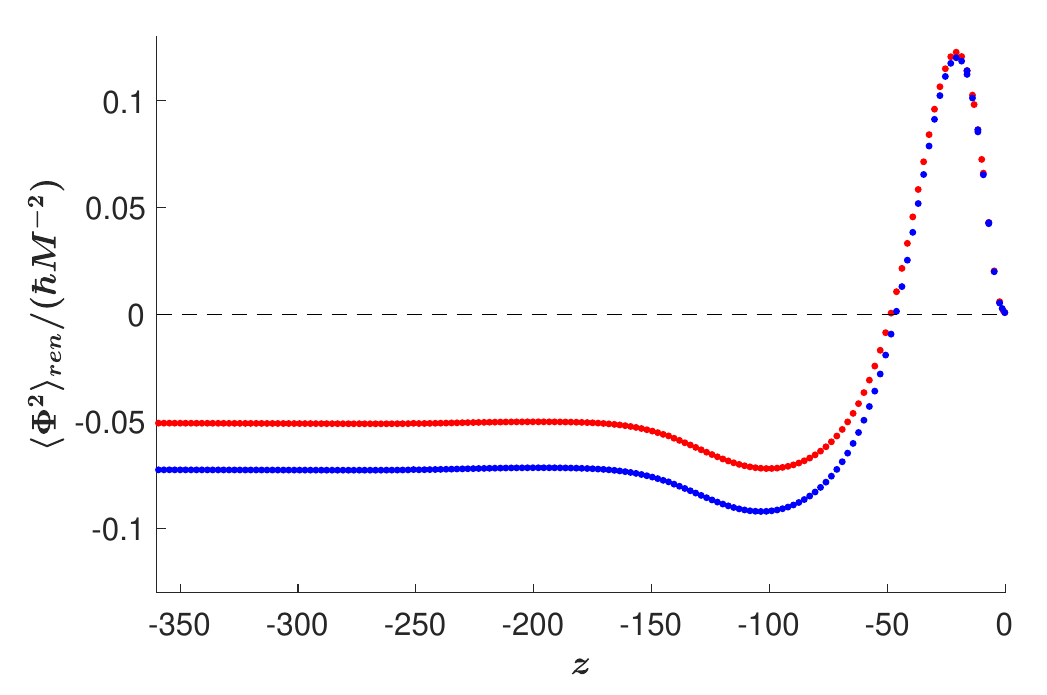}
\caption{$\langle\hat{\Phi}^{2}\left(r\right)\rangle_{ren}$ in the HH (red)
and Unruh (blue) states, as a function of $z$. }
\label{graph2-2} 
\end{figure}
 
\begin{figure}[h!]
\centering \includegraphics[width=0.45\textwidth,height=0.28\textheight]{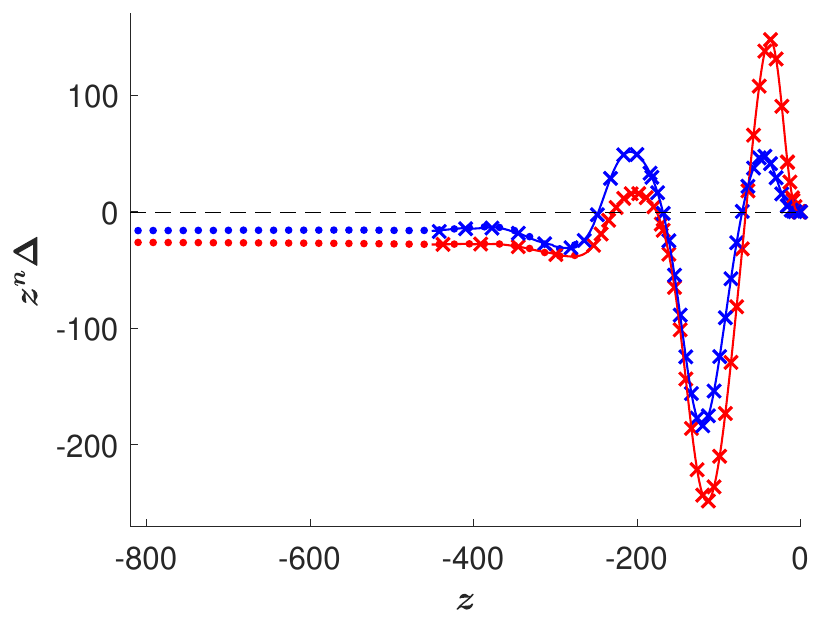}
\caption{$\Delta(z)\cdot z^{n}$ in the HH (red) and Unruh (blue) states, 
in the region $-845<z<0$ (which roughly corresponds to $10^{-367}<\delta r<1$).
The results for the Unruh state are divided here by a factor of $-150$,
for convenience. The plateaus at the left half of the $z$ axis indicate
the inverse-power behavior $\Delta\propto z^{-n}$. For both the HH
and Unruh results, the crosses indicate full numerical results, the
solid curves indicate semi-asymptotic results, and the dots indicate
the ``refined variant'' results. }
\label{graph2} 
\end{figure}

To our surprise, we found that this picture drastically changes once
we start exploring regions much closer to the IH. In fact, $\langle\hat{\Phi}^{2}\left(r\right)\rangle_{ren}$
eventually approaches a \emph{finite} value at $r\to r_{-}$, which
we  denote by  $(\hbar/M^{2})\langle\hat{\Phi}^{2}\rangle_{-}$
, where the index ``$-$'' refers to the limit $r\to r_{-}$. This
is clearly seen in Fig. 2, which displays $\langle\hat{\Phi}^{2}\rangle_{ren}$
as a function of the logarithmic variable $z$ defined by
\begin{equation}
z\equiv\ln(\delta r)\;,\qquad\delta r\equiv(r-r_{-})/M\:.\label{eq:z}
\end{equation}
Note that the IH corresponds to $z\to-\infty$. In both quantum states,
after a few quickly-decaying oscillations (there are actually two
maxima and two minima overall, although not all of them can be seen
in this figure),  $\langle\hat{\Phi}^{2}\rangle_{ren}$ approaches
a plateau. The asymptotic values are $\langle\hat{\Phi}^{2}\rangle_{-}^{H}\cong-0.05058$
and $\langle\hat{\Phi}^{2}\rangle_{-}^{U}\cong-0.07258$. Hereafter,
an index ``H'' or ``U'' will denote the HH state or Unruh state,
respectively.  

\paragraph*{Near-IH asymptotic behavior: }

To explore the near-IH asymptotic behavior we define (respectively
for each quantum state)
\begin{equation}
\Delta\equiv(M^{2}/\hbar)\langle\hat{\Phi}^{2}\rangle_{ren}-\langle\hat{\Phi}^{2}\rangle_{-}\;\label{eq:Delta}
\end{equation}
(i.e. the dimensionless deviation from $\langle\hat{\Phi}^{2}\rangle_{-}$). 

As it turns out, $\Delta(z)$ decays like $z^{-n}$, where hereafter
$n$ will stand for either $n_{H}=2$ (HH state) or $n_{U}=3$ (Unruh
state). To demonstrate this, Fig. 3 displays $z^{n}\cdot\Delta(z)$.
The flat horizontal forms of the red and blue lines, at the left half
of the $z$ axis, clearly indicate this leading-order behavior $\Delta\propto z^{-n}$
in the two quantum states. (This behavior is seen even more clearly
in Fig. 4.) 

The effective potential $V_{l}(r)$, given in (\ref{v}), vanishes
at the IH like $f\propto\delta r$. Therefore, for sufficiently small
$\delta r$ the radial equation (\ref{radeq}) becomes free, and its
general solution in that domain is
\begin{equation}
\psi_{\omega l}\cong A_{\omega l}e^{i\omega r_{*}}+B_{\omega l}e^{-i\omega r_{*}}\,.\qquad\quad(\delta r\ll1)\label{eq:Free}
\end{equation}
The coefficients $A_{\omega l},B_{\omega l}$ are dictated by the
scattering problem off the potential $V_{l}(r)$ from the EH to the
IH, and can be determined numerically. Note that on approaching the
IH $r_{*}$ diverges as $r_{*}\approx-z/2\kappa_{-}$, where $\kappa_{-}=(r_{+}-r_{-})/2r_{-}^{2}$
is the IH surface gravity. 

In order to explore the aforementioned inverse-power decay we need
to push the numerical solution to extremely small $\delta r$ values,
say $\delta r<e^{-400}\sim10^{-175}$, as can be seen in e.g. Fig.
3. This is hard to do with the brute-force numerical solution for
$\psi_{\omega l}$. {[}One of the difficulties, already seen in Eq.
(\ref{eq:Free}), is the very rapid variation of $\psi_{\omega l}$
with $\omega$ for $r_{*}\gg1$.{]} To overcome this difficulty, we
introduce the \emph{semi-asymptotic }approximation, in which we simply
employ Eq. (\ref{eq:Free}) as an approximation to $\psi_{\omega l}$
 for sufficiently small $\delta r$. The results obtained from this
approximation are displayed in Fig. 3 by the red and blue solid curves. 

Still, in the deep tails region (say $z<-700$) even this semi-asymptotic
approximation starts to be noisy (when numerically implemented to
explore the inverse-power tails). We therefore designed a refined
variant of this approximation, aimed to explore the tails region,
which can more efficiently take us to very large $|z|$ values. It
is this refined variant that we have used to produce Fig. 4 below
(and also the left region in Fig. 3). We point out that there are
nice overlap regions on the $z$ axis between these three slightly
different numerical procedures, as may be seen e.g. in Fig. 3. This
is further discussed in the Supplementary materials, which provide
additional information about the semi-asymptotic approximation and
its refined variant.

\paragraph*{Analytical expressions for the inverse-power tails: }

To our pleasant surprise, we found that it is possible to obtain,
analytically \cite{Maline}, the dominant inverse-power tails characterizing
the near-IH asymptotic behavior of $\Delta(z)$. This is possible
because, as it turns out, these tails are actually governed by the
small-$\omega$ asymptotic behavior of $A_{\omega l}$, $B_{\omega l}$,
and $\rho_{\omega l}^{\mathrm{up}}$; and this small-$\omega$ behavior
can be deduced analytically. This analysis yields the two dominant
inverse powers ($n_{H}=2$ and $n_{U}=3$) as well as their multiplicative
amplitude parameters (for both quantum states). 

Furthermore, since we had to carry the analysis to order $z^{-3}$
(needed for the Unruh-state leading order), we actually got, almost
for free, the term $\propto z^{-3}$ for the HH state as well. Thus,
including all the inverse-power terms to which we presently have analytical
access, we write the tail expressions as
\begin{equation}
\Delta_{U}=C_{U}z^{-3}+...\;,\;\;\Delta_{H}=C_{H}z^{-2}+C_{H}^{1}z^{-3}+...\;,\label{eq:HUtails}
\end{equation}
where ``...'' denotes higher-order corrections. Defining $\alpha\equiv r_{+}/r_{-}$,
we find 
\begin{equation}
C_{U}=2\Lambda\left(1-\alpha^{4}\right)\left(1-\alpha\right)^{2}\left(11+14\alpha+11\alpha^{2}\right)\label{eq:CU-1-1}
\end{equation}
\begin{equation}
C_{H}=3\Lambda\,\alpha^{-2}\left(1-\alpha^{4}\right)^{2}\label{eq:CH}
\end{equation}
\begin{equation}
C_{H}^{1}=2\log\left[\frac{2\left(\alpha-1\right)}{\alpha+1}\right]C_{H}-\frac{1}{4}\left(\alpha^{-2}-3\right)C_{U}\label{eq:CH1}
\end{equation}
where $\Lambda\equiv\left(1-\alpha^{2}\right)/768\pi^{2}$.

Figure 4 displays the analytical expressions (\ref{eq:HUtails}) (black
curves) and the numerical data (dots) for the inverse-power tails,
for both quantum states, in the range $400<-z<1500$. It shows excellent
agreement, supporting the validity/accuracy of both the theoretical
analysis and numerics. 
\begin{figure}[h!]
\centering \includegraphics[width=0.45\textwidth,height=0.28\textheight]{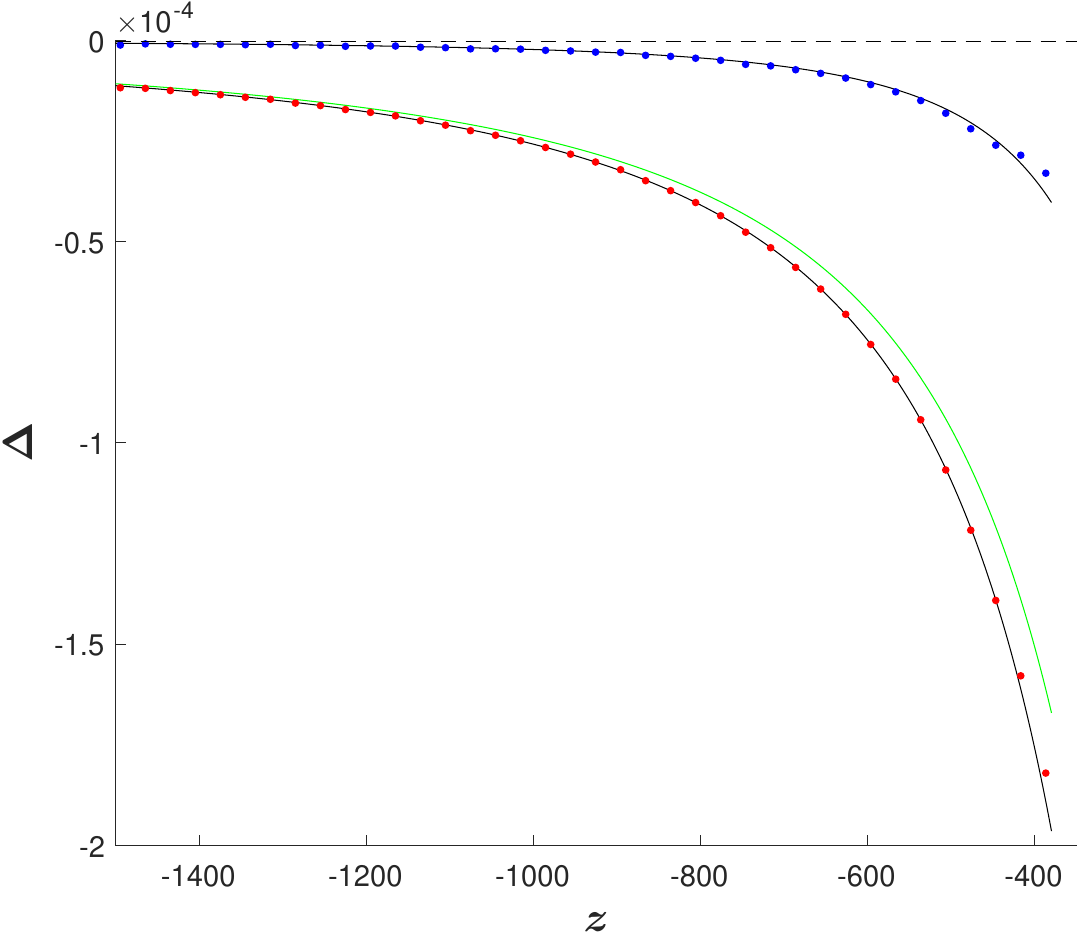}
\caption{$\Delta(z)$ in the HH state (red dots) and Unruh state (blue dots)
exceedingly close to $r_{-}$ (up to $z\sim1500$, which roughly corresponds
to $\delta r\sim10^{-650}$), computed using the ``refined variant''.
The black curves are the analytical expressions (\ref{eq:HUtails})
for the inverse-power tails.  The green curve indicates the leading-order
analytical result $\Delta_{H}\approx C_{H}z^{-2}$ (whereas the corresponding
black curve also includes the next-order term $C_{H}^{1}z^{-3}$).}
\label{graph4} 
\end{figure}

\paragraph*{Trace of the stress tensor: }

For a minimally coupled massless scalar field, the RSET trace $\langle\hat{T}_{\mu}^{\mu}\rangle_{ren}$
is uniquely determined \cite{Group:2018} by $\langle\hat{\Phi}^{2}(x)\rangle_{ren}$
via 
\begin{equation}
\langle\hat{T}_{\mu}^{\mu}\rangle_{ren}=-\frac{1}{2}\square\langle\hat{\Phi}^{2}(x)\rangle_{ren}+(\mbox{local term}).\label{eq:trace}
\end{equation}
The local term only depends on the background metric, which is perfectly
regular at the IH. Therefore the singular piece of $\langle\hat{T}_{\mu}^{\mu}\rangle_{ren}$
is fully described by the D'Alembertian term.  Since the constant
$\langle\hat{\Phi}^{2}\rangle_{-}$ contributes nothing to the D'Alembertian,
we are left with $-(\hbar/2M^{2})\square\Delta$. Applying the D'Alembertian
operator to Eq. (\ref{eq:HUtails}), we obtain for the two quantum
states, at leading order in $1/z$ (and $\delta r$): 
\begin{equation}
\langle\hat{T}_{\mu}^{\mu}\rangle_{ren}\cong n(n+1)\frac{\hbar}{M^{2}}\,\kappa_{-}\,\frac{C}{r-r_{-}}\,z^{-n-2}\:,\label{eq:Trace_Result}
\end{equation}
where, recall, $n_{H}=2$, $n_{U}=3$ and $C$ is either $C_{H}$
or $C_{U}$ specified above. 

\paragraph*{Discussion: }

We found that $\langle\hat{\Phi}^{2}\rangle_{ren}$ is finite at the
IH. This finite asymptotic value is approached via a few quickly decaying
oscillations followed by an inverse-power tail. In turn, the RSET
trace $\langle\hat{T}_{\mu}^{\mu}\rangle_{ren}$ diverges as $1/(r-r_{-})$
softened by a certain inverse power of $\ln(r-r_{-})$. We obtained
a fully analytical description of this divergent trace (at leading
order), Eq. (\ref{eq:Trace_Result}).

Here we only investigated numerically the case $Q/M=0.8$. However,
our results for the inverse power tails \textemdash{} and, more importantly,
for the asymptotic divergence (\ref{eq:Trace_Result}) of the RSET
trace \textemdash{} apply to any (non-extremal) $M$ and $Q$. 

The behavior of $\langle\hat{\Phi}^{2}\rangle_{ren}$ on approaching
the IH is remarkably complex. In particular, the final inverse-power
tails are only exposed at, say, $\delta r<10^{-175}$. This complex
asymptotic behavior may be traced to the factors $e^{\pm i\omega r_{*}}$
in Eq. (\ref{eq:Free}). The mode contribution to $\langle\hat{\Phi}^{2}\rangle_{ren}$
contains terms quadratic in $\psi_{\omega l}$, including factors
$e^{\pm2i\omega r_{*}}$ ( multiplying certain functions of $A(\omega),B(\omega)$,
etc.). Integration over $\omega$ then leaves a nontrivial function
of $r_{*}$, embodied in the asymptotic behavior of $\langle\hat{\Phi}^{2}\rangle_{ren}$.

It is interesting to compare these results to a recent work \cite{RNOrr:2018}
carried out by one of us (OS), in which the large-$l$ approximation
was used to obtain bounds on the divergence rate of $\langle\hat{\Phi}^{2}\rangle_{ren}$,
$\langle\hat{T}_{\mu}^{\mu}\rangle_{ren}$, and certain components
of $\langle\hat{T}_{\mu\nu}\rangle_{ren}$. In particular it was found
that for both the Unruh and HH states $\langle\hat{\Phi}^{2}\rangle_{ren}$
and $\langle\hat{T}_{\mu}^{\mu}\rangle_{ren}$ must be less divergent
than $1/(r-r_{-})$ and $1/(r-r_{-})^{2}$, respectively. The results
presented here for these two quantities are fully consistent with
these bounds.

The expressions presented here for the pre-factors $C_{H},C_{U}$
that control the divergence of $\langle\hat{T}_{\mu}^{\mu}\rangle_{ren}$,
only apply to a minimally-coupled massless scalar field. In the case
of non-minimal coupling they will change. In particular, in the case
of conformal coupling these pre-factors will vanish altogether, because
the standard trace-anomaly formula guarantees regularity of the trace
at the IH. The same situation will occur in the case of a quantum
electromagnetic field, since this field is conformal too. 

It is still unclear, however, if the \emph{gravitational} semiclassical
contribution to the effective stress-energy will possess such a trace
divergence at the IH. The presence of a gravitational contribution
(associated with quantized linearized modes of the gravitational field)
to the effective $\langle\hat{T}_{\mu\nu}\rangle_{ren}$ is obvious
from the very basic fact that gravitons do significantly contribute
to Hawking radiation \cite{Don} (and correspondingly, negative semiclassical
gravitational-field influx must penetrate into the EH of the evaporating
BH and contribute to its shrinkage). However, a formalism for quantifying
the semiclassical effective gravitational stress-energy tensor has
not been formulated so far.

This analysis calls for extension in several obvious directions. The
first obvious step is to elevate the analysis from $\langle\hat{\Phi}^{2}\rangle_{ren}$
to the RSET. Second, the quantum scalar field should better be replaced
by the (more realistic) quantum electromagnetic field. In addition,
it will be important to extend the analysis from RN to the Kerr background
(a spinning BH), which is obviously much more realistic than a spherical
charged BH. 

Finally, it will be very interesting (but also very challenging) to
explore the back-reaction effect of the semiclassical RSET on the
BH interior, according to the semiclassical Einstein equation (\ref{semi_ein}). 

\

\begin{acknowledgements} We thank Robert Wald, Marc Casals,
	and Adrian Ottewill for interesting and helpful discussions. This
	research was supported by the Asher Fund for Space Research at the
	Technion. The work of AL and OS was further supported by the Israel
	Science Foundation under Grant No. 1696/15 and by the I-CORE Program
	of the Planning and Budgeting Committee. The work of NZ was partly
	supported by the Israel Science Foundation under Grant No. 600/18.
\end{acknowledgements}

\section*{Supplemental Material}

\subsection{Method of computation of{\normalsize{} $\langle\hat{\Phi}^{2}\rangle_{ren}$
	}\label{subsec:Full_scheme} }

We use here the mode-sum expression derived in Ref. {[}26{]} for the
TPF, to which we apply the $\theta$-splitting variant {[}19{]} of
the PMR method. In this treatment, we follow the same procedure used
in Sec. 3 of Ref. {[}27{]}, where a step-by-step recipe for the computation
of $\langle\hat{\Phi}^{2}\rangle_{ren}$ inside a Schwarzschild BH
is provided. Here we use exactly the same procedure except for the
following trivial modifications (which follow directly from the change
of the metric from Schwarzschild to RN): (i) The EH surface gravity,
$\kappa$ of Ref. {[}27{]} , is here replaced by $\kappa_{+}\equiv\left(r_{+}-r_{-}\right)/2r_{+}^{2}$.
(ii) $d(r)$ of Eq. (3.11) therein is here replaced by $d\left(r\right)=\left(Q^{2}-Mr\right)/24\pi^{2}r^{4}$.
(iii) The counter-term $G_{DS}(x,x')$, described in Eq. (3.3) therein
for the Schwarzschild case, now has a (finite) additional term proportional
to the Ricci tensor. Note, however, that this change is already incorporated
in the aforementioned change in $d(r)$. Therefore, in the operational
part of Sec. 3 of Ref. {[}27{]} only changes (i) and (ii) need be
considered.  

\subsection{Numerical parameters}

In the purpose of computing $\langle\hat{\Phi}^{2}\rangle_{ren}$
in the vicinity of the IH, the radial equation was solved for 11 $l$
values ($0\leq l\leq10$), and for each $l$ in the range $\omega\in\left[0,10M^{-1}\right]$
with uniform spacings of $d\omega=0.0005M^{-1}$ and $d\omega=0.004M^{-1}$
in the regions $\omega\in\left[0,2M^{-1}\right]$ and $\omega\in\left[2M^{-1},10M^{-1}\right]$,
respectively. The computation of $\langle\hat{\Phi}^{2}\rangle_{ren}$
in the general region between the EH and IH (namely $0.5\leq r\leq1.6$)
involved solving the radial equation for 31 $l$ values ($0\leq l\leq30$),
and for each $l$ in the range $\omega\in\left[0,120M^{-1}\right]$
with uniform spacings of $d\omega=0.0005M^{-1}$ and $d\omega=0.1M^{-1}$
in the regions $\omega\in\left[0,6M^{-1}\right]$ and $\omega\in\left[6M^{-1},120M^{-1}\right]$,
respectively. For the computation in the region close to the EH, the
radial equation was solved for 16 $l$ values ($0\leq l\leq15$),
and for each $l$ in the range $\omega\in\left[0,15M^{-1}\right]$
with uniform spacings of $d\omega=0.01M^{-1}$. 

\subsection{The semi-asymptotic approximation and its refined variant}

A key ingredient in the computation of $\langle\hat{\Phi}^{2}\rangle_{ren}$
inside the BH is the quantity $E_{\omega l}(r)$, given in Eqs. (3.7,
3.8) of Ref. {[}27{]} (for the Unruh and HH states respectively).
It involves the function $\bar{\psi}_{\omega l}$ which is specified
in Eq. (3.5) therein in terms of the radial function $\psi_{\omega l}$.

In the semi-asymptotic approximation we leave $E_{\omega l}(r)$ unchanged
except that we substitute the near-IH asymptotic expression (9) for
$\psi_{\omega l}$, and also substitute $r_{-}$ for $r$. For the
HH state we get: 
\begin{widetext}
	\begin{equation}
	E_{\omega l}^{H}=\frac{1}{2\pi|\omega|r_{-}^{2}}\mathrm{Re}\left[\left(\left|A_{\omega l}\right|^{2}+\left|B_{\omega l}\right|^{2}+2A_{\omega l}B_{\omega l}^{*}e^{2i\omega r_{*}}\right)\coth\tilde{\omega}+\frac{\rho_{\omega l}^{\mathrm{up}}}{\sinh\tilde{\omega}}\left(A_{\omega l}^{2}e^{2i\omega r_{*}}+B_{\omega l}^{2}e^{-2i\omega r_{*}}+2A_{\omega l}B_{\omega l}\right)\right]\label{eq:Efull}
	\end{equation}
\end{widetext}

where $\tilde{\omega}\equiv\pi\omega/\kappa_{+}$. In principle, this
quantity needs be subsequently integrated over $\omega$ (and also
summed over $l$). Note, however, that at large $\omega$ both $A_{\omega l}$
and $\rho_{\omega l}^{\mathrm{up}}$ decay exponentially while $B_{\omega l}\to1$,
hence $E_{\omega l}^{H}\propto\omega^{-1}$ and its integral over
$\omega$ diverges. To this end, in the original $\theta$-splitting
method we subtract $E_{\omega,l=0}$ from $E_{\omega l}$ before integration,
see e.g. Eq. (3.9) therein. This $l=0$ subtraction is also done in
the semi-asymptotic approximation. 

In the refined variant we proceed in a different manner. From the
squared brackets in Eq. (\ref{eq:Efull}) we subtract two terms: 
\begin{equation}
\delta E_{1}\equiv\left(\left|A_{\omega l}\right|^{2}+\left|B_{\omega l}\right|^{2}\right)\coth\tilde{\omega}+\frac{2\rho_{\omega l}^{\mathrm{up}}}{\sinh\tilde{\omega}}\,A_{\omega l}B_{\omega l}\label{r indep}
\end{equation}
and 
\begin{equation}
\delta E_{2}\equiv\frac{1}{\sinh\tilde{\omega}}\left(2A_{0}B_{0}^{*}-A_{0}^{2}e^{2i\omega r_{*}}-B_{0}^{2}e^{-2i\omega r_{*}}\right),\label{l independ}
\end{equation}
where $A_{0}$ and $B_{0}$ denote the $\omega\to0$ limits of $A$
and $B$ respectively. {[}These parameters can actually be obtained
analytically, they are given by $(-1)^{l}(r_{-}^{2}\pm r_{+}^{2})/2r_{+}r_{-}$,
with the $+$ sign for $B_{0}$ and the $-$ sign for $A_{0}$, but
we shall not discuss this derivation here.{]} The subtraction of $\delta E_{1}$
removes the $\propto\omega^{-1}$ term at large $\omega$. In turn,
the subtraction of $\delta E_{2}$ removes the $\omega\to0$ divergence
that would have been caused by the $\delta E_{1}$ subtraction. We
denote the modified $E_{\omega l}^{H}$ (due to this subtraction)
by $\hat{E}_{\omega l}^{H}$ for later convenience.

Obviously, this subtraction of $\delta E_{1}$ and $\delta E_{2}$
needs to be justified. The subtraction of $\delta E_{2}$ is allowed
because it is independent of $l$, and adding any $l$-independent
term to the mode contribution $E_{\omega l}$ does not affect the
resultant $\langle\hat{\Phi}^{2}\rangle_{ren}$ at all (it just modifies
the ``blind spot'' {[}19{]}). 

The other term $\delta E_{1}$ does depend on $l$, but nevertheless
it is independent of $r_{*}$. As a consequence, although $\langle\hat{\Phi}^{2}\rangle_{ren}$
is changed by this subtraction, this change merely amounts to adding
some $r$-independent constant to $\langle\hat{\Phi}^{2}\rangle_{ren}$.
Therefore, this subtraction does not affect $\Delta(z)$ at all. We
can thus use the refined version in e.g. Figs. 3 and 4, which describe
$\Delta(z)$ (but not in Figs. 1 and 2 that describe the full $\langle\hat{\Phi}^{2}\rangle_{ren}$;
see also last paragraph). 

Since the subtraction of $\delta E_{1}$ removes the large-$\omega$
irregularity, we no longer need to subtract the $l=0$ contribution.
Thus, in Eq. (3.9) of Ref. {[}27{]} the integrand (i.e. the term in
squared bracket) is now simply replaced by $\hat{E}_{\omega l}^{H}$.
This new integrand actually decays (at large $\omega$) \emph{exponentially}
in $\omega$, leading to a quick and efficient numerical convergence
of the integral. We denote the resultant integral by $\hat{F}(l,r)$. 

We find, somewhat surprisingly, that the quantity $\hat{F}_{reg}(l,r)=\hat{F}(l,r)-F_{sing}(l,r)$
\emph{decays to zero} at large $l$ (i.e. there is no blind spot).
Therefore, in the refined variant there is no need to define the function
$H$ {[}Eq. (3.14) therein{]}. Instead, the final result is simply
obtained (up to some $r$-independent shift) by summing $[(2l+1)/(4\pi)]\hat{F}_{reg}(l,r)$
over $l$. 

Furthermore, at large $-z$ this sum over $l$ converges tremendously
fast. For example, the numerical data presented in Fig. 4 were obtained
by summing over $l$ up to $l=2$. But in fact, the contribution from
$l=0$ looks just the same; the contribution from $l=1$ cannot be
seen in the graph (let alone the $l=2$ contribution). Although unnecessary,
we chose to include the $l=1$ and $l=2$ terms as well in our computation. 

So far we focused on the HH state for concreteness. The refined computation
of $\Delta$ in the Unruh state may proceed in a fully analogous manner.
In both quantum states, the exponential decay of the integrand with
$\omega$, combined with the extremely fast convergence of the sum
of $\hat{F}_{reg}(l)$ over $l$, lead  to numerical results of a
considerably better quality, which is the purpose we sought to fulfill
when we employed the refined method. 

Finally, we use this opportunity to clarify again which of the three
variants was used in each of the figures. Both Figs. 1 and 2 were
produced by the full-fledged computational scheme. In contrast, Fig.
4 displays the refined results for $\Delta(z)$. Figure 3, which presents
$z^{n}\cdot\Delta(z)$, displays the results obtained from all three
variants.  Notice the ranges of overlaps between these different
variants: The full-fledged numerics and the semi-asymptotic approximation
overlap in the very wide domain $-450<z<-3$. In addition, all three
variants overlap in the range $-450<z<-280$. This figure demonstrates
nice agreement between the different variants throughout the domains
of their overlap.


\noindent 

\noindent 

\begin{thebibliography}{10}
\bibitem{Hiscock:1981} W.~A.~Hiscock, Evolution of the interior
of a charged black hole, Phys.\ Rev.\ Lett.\ \textbf{83}, 110 (1981).

\bibitem{PoissonIsrael:1990} E.~Poisson and W.~Israel, Internal
structure of black holes, Phys.\ Rev.\ D.\ \textbf{41}, 1796 (1990).

\bibitem{OriMassInflation:1991} A.~Ori, Inner structure of a charged
black hole: An exact mass-inflation solution, Phys.\ Rev.\ Lett.\ \textbf{67},
789 (1991).

\bibitem{BradySmith:1995} P.~R.~Brady and J.~D.~Smith, Black
hole singularities: a numerical approach, Phys.\ Rev.\ Lett.\ \textbf{75},
1256 (1995).

\bibitem{Piran}S. Hod and T. Piran, Mass Inflation in Dynamical Gravitational
Collapse of a Charged Scalar Field, Phys. Rev. Lett. 81, 1554 (1998).

\bibitem{Burko:1997} L.~M.~Burko, Structure of the black hole's
Cauchy-horizon singularity, Phys.\ Rev.\ Lett.\ \textbf{79}, 4958
(1997).

\bibitem{Ori:1992} A.~Ori, Structure of the singularity inside a
realistic rotating black hole, Phys.\ Rev.\ Lett.\ \textbf{68},
2117 (1992).

\bibitem{BradyDrozMorsnik:1998} P.~R.~Brady, S.~Droz, and S.~M.~Morsnik,
Late-time singularity inside nonspherical black holes, Phys.\ Rev.\ D.\ \textbf{58},
084034 (1998).

\bibitem{Ori:1999} A.~Ori, Oscillatory null singularity inside realistic
spinning black holes, Phys.\ Rev.\ Lett.\ \textbf{83}, 2117 (1999).

\bibitem{Tipler}F. J. Tipler, Singularities in conformally flat spacetimes,
Phys. Lett. A 64, 8 (1977).

\bibitem{Ori-weak}A. Ori, Strength of curvature singularities, Phys.
Rev. D 61, 064016 (2000).

\bibitem{Dafermos:2017} M.~Dafermos and J.~Luk, The interior of
dynamical vacuum black holes I: The $C^{0}$-stability of the Kerr
Cauchy horizon, {[}arXiv:1710.01722 {[}gr-qc{]}{]} (2017).

\bibitem{BirrellDavies:1978} N.~D.~Birrell and P.~C.~W.~Davies,
On falling through a black hole into another universe, Nature (London)
\textbf{272}, 35 (1978).

\bibitem{Ottewill:2000} A.~C.~Ottewill and E.~Winstanley, Renormalized
stress tensor in Kerr space-time: General results, Phys.\ Rev.\ D.\ \textbf{62},
084018 (2000).

\bibitem{Hiscock:1980} W.~A.~Hiscock, Quantum-mechanical instability
of the Kerr-Newman black-hole interior, Phys.\ Rev.\ D.\ \textbf{21},
2057 (1980).

\bibitem{Hawking:1974} S.~W.~Hawking, Black hole explosions?, Nature
\textbf{248}, 30 (1974).

\bibitem{Hawking:1975} S.~W.~Hawking, Particle creation by black
holes, Commun.\ Math.\ Phys.\ \textbf{43}, 199 (1975).

\bibitem{AAt:2015} A.~Levi and A.~Ori, Pragmatic mode-sum regularization
method for semiclassical black-hole spacetimes, Phys.\ Rev.\ D.\ \textbf{91},
104028 (2015).

\bibitem{AAtheta:2016} A.~Levi and A.~Ori, Mode-sum regularization
of $\langle{\phi}^{2}\rangle$ in the angular-splitting method, Phys.\ Rev.\ D.\ \textbf{94},
044054 (2016).

\bibitem{Christensen:1976} S.~M.~Christensen, Vacuum expectation
value of the stress tensor in an arbitrary curved background: The
covariant point separation method, Phys.\ Rev.\ D.\ \textbf{14},
2490 (1976).

\bibitem{Christensen:1978} S.~M.~Christensen, Regularization, renormalization,
and covariant geodesic point separation, Phys.\ Rev.\ D.\ \textbf{17},
946 (1978).

\bibitem{Ander_His_Sam:1995} P.~R.~Anderson, W.~A.~Hiscock and
D.~A.~Samuel, Stress-energy tensor of quantized scalar fields in
static spherically symmetric spacetimes, Phys.\ Rev.\ D.\ \textbf{51},
4337 (1995).

\bibitem{Unruh:1976} W.~G.~Unruh Notes on black-hole evaporation,
Phys.\ Rev.\ D.\ \textbf{14}, 870 (1976).

\bibitem{HH:1976} J.~B.~Hartle and S.~W.~Hawking Path-integral
derivation of black-hole radiance, Phys.\ Rev.\ D.\ \textbf{13},
2188 (1976).

\bibitem{Israel:1976} W.~Israel Thermo-field dynamics of black holes,
Phys.\ Lett.\ A.\ \textbf{57}, 107 (1976).

\bibitem{Group:2018} A.~Lanir, A.~Levi, A.~Ori and O.~Sela Two-point
function of a quantum scalar field in the interior region of a Reissner-Nordstrom
black hole, Phys.\ Rev.\ D.\ \textbf{97}, 024033 (2018).

\bibitem{SchAssaf:2018} A.~Lanir, A.~Levi, and A.~Ori, Mode-sum
renormalization of $\langle\hat{\Phi}^{2}\rangle$ for a quantum scalar
field inside a Schwarzschild black hole, Phys.\ Rev.\ D.\ \textbf{98},
084017 (2018).

\bibitem{Frolov}V. P. Frolov, Vacuum polarization near the event
horizon of a charged rotating black hole, Phys.\ Rev.\ D.\ 26,
954 (1982).

\bibitem{Ottewill}A. Ottewill, private communication.

\bibitem{Maline}A. Maline et al (in preparation).

\bibitem{RNOrr:2018} O.~Sela, Quantum effects near the Cauchy horizon
of a Reissner-Nordstrom black hole, Phys.\ Rev.\ D.\ \textbf{98},
024025 (2018). 

\bibitem{Don}D. N. Page, Particle emission rates from a black hole:
Massless particles from an uncharged, nonrotating hole, Phys.\ Rev.\ D.\ \textbf{13},
198 (1976).

\bibitem{Foot1}The results described below for $\langle\hat{\Phi}^{2}\rangle_{ren}$
are actually independent of the coupling constant (because the Ricci
scalar vanishes in our case). However, the results for $\langle\hat{T}_{\mu}^{\mu}\rangle_{ren}$
only apply for a minimally-coupled field.

\bibitem{Foot1a}We shall consider here a fixed RN geometry, without
back-reaction.

\bibitem{Foot2}Throughout, by ``interior region'' we actually refer
to the 'predictable' internal domain $r_{-}\leq r\leq r_{+}$.
\end{thebibliography}
\end{document}